\documentclass{iau_FM}
\usepackage{graphicx}

\title[Star formation in galaxy clusters] %% give here short title %%
{The spatially resolved view of star formation \\ in galaxy clusters}

\author[Bianca M. Poggianti \& the GASP team]   %% give here short author list %%
{Bianca M. Poggianti$^1$
 \and the GASP team}

\affiliation{$^1$INAF-Osservatorio Astronomico di Padova, \\ vicolo dell'Osservatorio 5, 35122 Padova, Italy
 \\ email: {\tt bianca.poggianti@inaf.it}}

\begin{document}

\maketitle

\begin{abstract}
Integral field spectroscopic studies of galaxies in dense
environments, such as clusters and groups of galaxies, have provided
new insights for understanding how star formation proceeds, and
quenches. I present the spatially resolved view of the star
formation activity and its link with the multiphase gas in cluster galaxies
based on MUSE
and multi-wavelength data of the GASP survey.
I discuss the link among the different
scales (i.e. the link between the spatially resolved and the global star
formation rate-stellar mass relation), the spatially
resolved signatures and
the quenching histories of jellyfish (progenitors) and
post-starburst (descendants) galaxies in clusters.
Finally, I discuss the multi-wavelength view of
star-forming clumps both in galaxy disks and in the
tails of stripped gas.
%Integral field spectroscopic studies of galaxies in dense
%environments, such as clusters and groups of galaxies, have provided
%new insights for understanding how star formation proceeds, and
%quenches. I will present the spatially resolved view of the star
%formation activity and its link with the multiphase gas based on MUSE
%and multi-wavelength data of the GASP survey, discussing the relation
%among the different scales (eg the link between the spatially resolved
%and the global star formation rate-stellar mass relation). I will
%present the spatially resolved signatures of different quenching
%physical mechanisms, and the detailed quenching histories of jellyfish
%(progenitors) and post-starburst (descendants) galaxies in clusters at
%z~0 and at z~0.4. I will show the combined HST, MUSE and ALMA view of
%star-forming clumps in disks but also those forming in situ in the
%tails of gas stripped by ram pressure stripping, presenting their
%sizes, stellar ages, masses, scaling relations at different
%wavelengths. Finally, I will show evidence for internally driven
%quenching due to AGN activity, and how the latter can be connected
%with external processes such as ram pressure.
\end{abstract}

\firstsection % if your document starts with a section,
\section{Introduction}

When galaxies infall into clusters and move at high speed with respect
to the hot intracluster medium (ICM), the ram pressure (RP) exerted by
the ICM upon the galaxy interstellar medium (ISM) can remove the
galaxy gas, wherever the RP exceeds the gravitation pull (Gunn \& Gott
1972, Jaffe' et al. 2018).  The hot halo gas is affected first, then the disk gas is
removed starting from the outer disk regions and stripping works its
way inward in the disk, until the galaxy remains eventually devoid of gas.

\begin{figure}[b]
% \vspace*{-2.0 cm}
\begin{center}
 \includegraphics[width=5.5in]{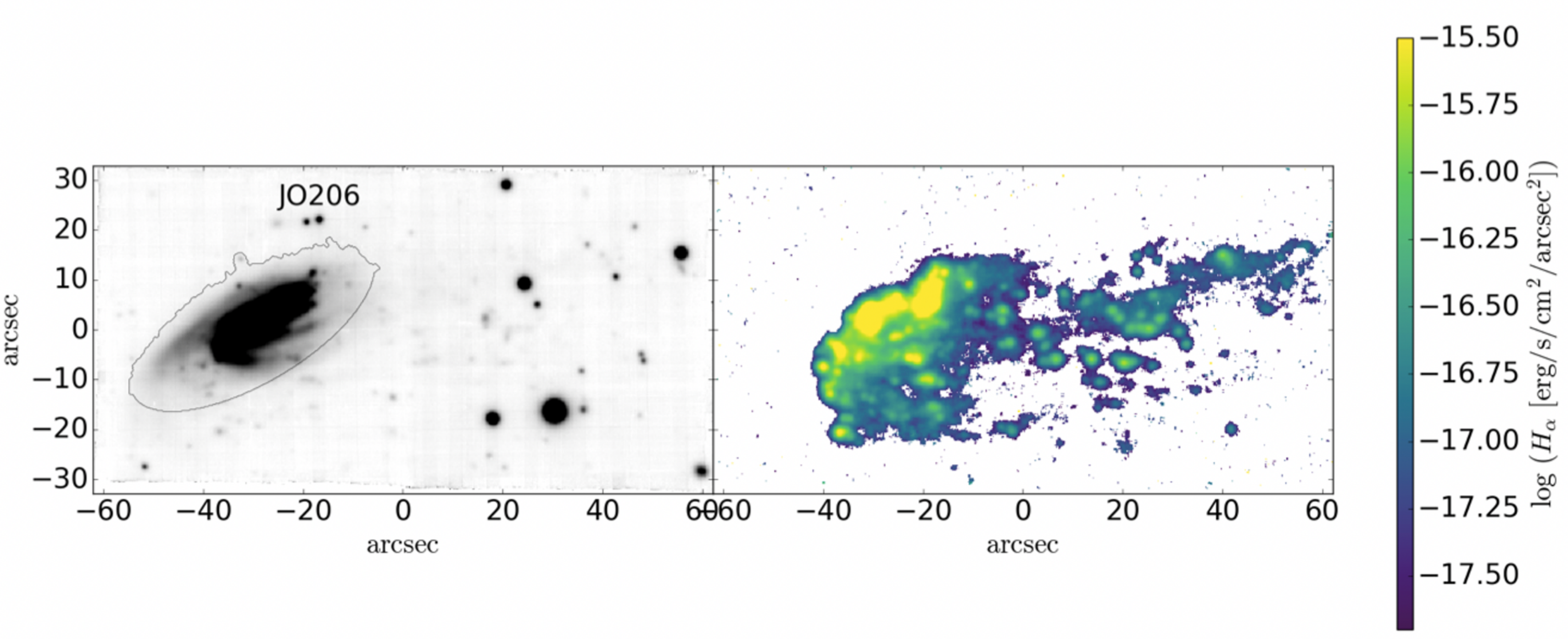}
% \vspace*{-1.0 cm}
\caption{The jellyfish galaxy JO206, a massive ($9 \times 10^{10} M_{\odot}$) galaxy in a low mass cluster (velocity dispersion $\sim 500 \rm km \, s^{-1}$). Left. White-light MUSE image. Right. MUSE map of $\rm H\alpha$ emission. Note the tentacles of stripped ionized gas departing for $\sim 80$ kpc from the disk, to the west. Clearly visible in this tail are $\rm H\alpha$-emitting, star-forming clumps. From Poggianti et al. (2017a).} 
   \label{fig1}
\end{center}
\end{figure}

\begin{figure}[b]
% \vspace*{-2.0 cm}
\begin{center}
 \includegraphics[width=3.2in]{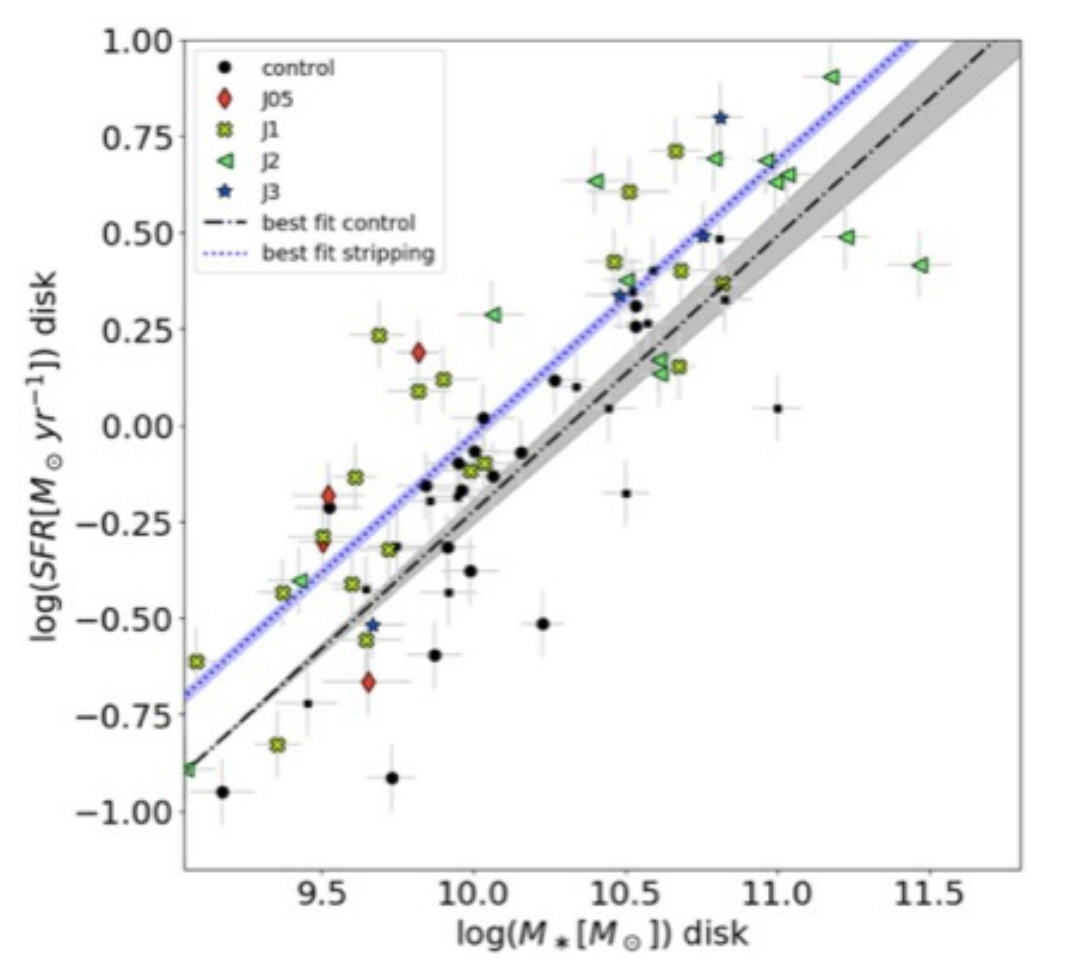}
% \vspace*{-1.0 cm}
 \caption{Global SFR-M relation for undisturbed galaxies (black points, grey line fit) versus RP stripped galaxies (colored points, light blue line fit). RP galaxies show a modest but significant enhancement in their SF activity. From Vulcani et al. (2018)}
   \label{fig1}
\end{center}
\end{figure}

%\begin{figure}[b]
%% \vspace*{-2.0 cm}
%\begin{center}
%\includegraphics[width=3.0in]{fig3.pdf}
%% \vspace*{-1.0 cm}
% \caption{Characteristics of a sample of PSB galaxies in z=0.3-0.4 clusters. Top. Time since quenching. Bottom. Fraction of mass assembled during the last 1.5 Gyr, used to roughly estimate the mass fraction involved in the burst prior to truncation. Boxes represent the 25th and 75th percentiles, the median is the horizontal line and the whiskers are the 15th and 85th percentiles. From Werle et al. (2022).}
%   \label{fig1}
%\end{center}
%\end{figure}

\begin{figure}[b]
%\vspace*{-4.0 cm}
\begin{center}
\vspace*{-5.0cm}
 \includegraphics[width=5.0in]{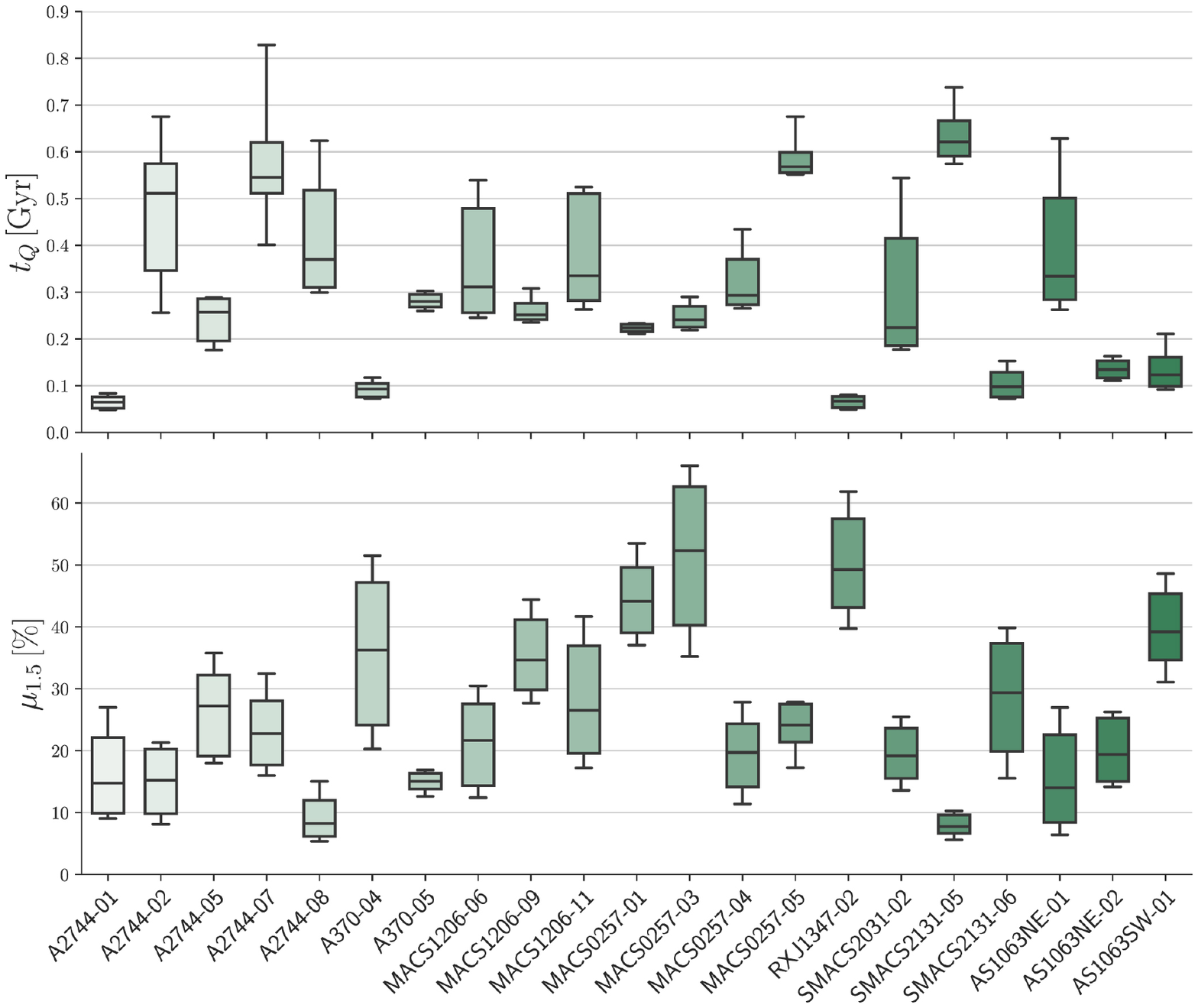}
\vspace*{-4.3 cm}
 \caption{Characteristics of a sample of PSB galaxies in z=0.3-0.4 clusters. Top. Time since quenching. Bottom. Fraction of mass assembled during the last 1.5 Gyr, used to roughly estimate the mass fraction involved in the burst prior to truncation. Boxes represent the 25th and 75th percentiles, the median is the horizontal line and the whiskers are the 15th and 85th percentiles. From Werle et al. (2022).}
%\caption{Maps of the time elapsed since quenching for PSB galaxies in distant clusters. Several cases of outside-in quenching are visible, as well as some cases for side-to-side quanching and inside-out (AGN) quenching.From Werle et al. (2022).}
   \label{fig1}
\end{center}
\end{figure}

\begin{figure}[b]
% \vspace*{-2.0 cm}
\begin{center}
\includegraphics[width=5.6in]{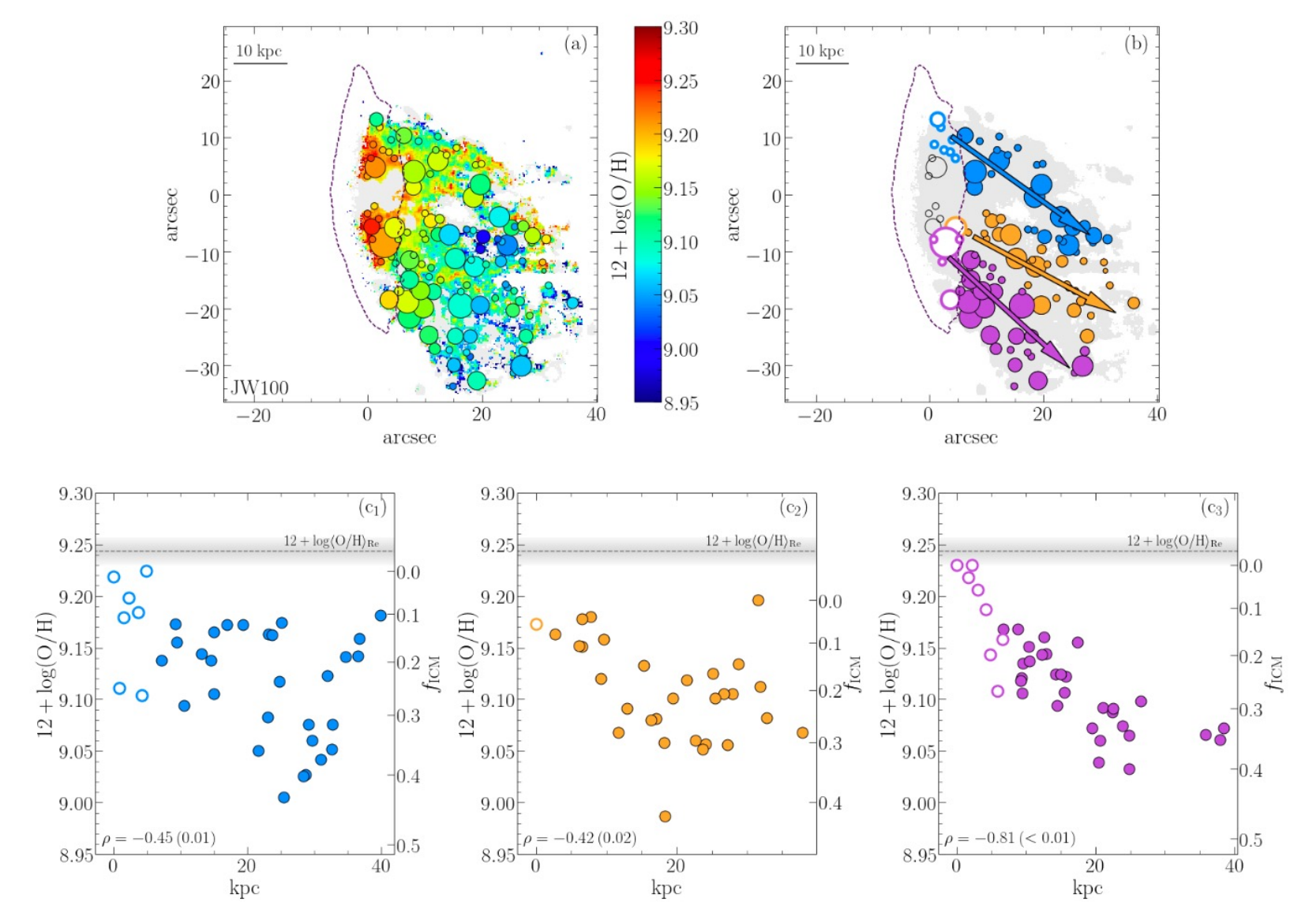}
% \vspace*{-1.0 cm}
 \caption{MUSE gas metallicity map of the galaxy JW100 (top left). Three main tails can be identified (top right). The metallicity strongly decreases along each tail (bottom panels). Assuming a typical ICM metallicity of 0.3 times solar, the fraction of gas provided by ICM mixing can be estimated (right Y-axes in bottom panels). From Franchetto et al. (2021).}
   \label{fig1}
\end{center}
\end{figure}

\begin{figure}[b]
% \vspace*{-2.0 cm}
\begin{center}
\includegraphics[width=3.4in]{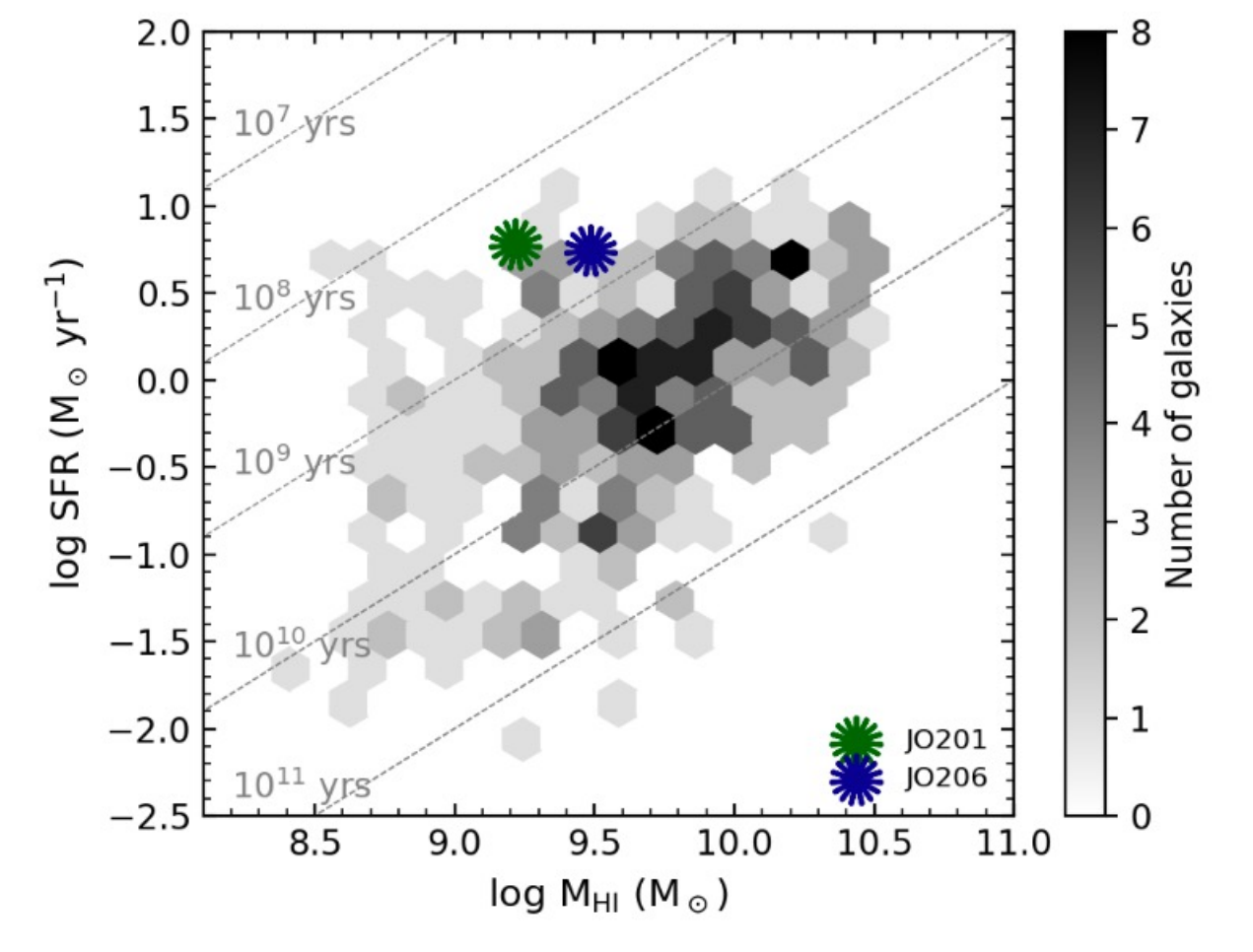}
% \vspace*{-1.0 cm}
 \caption{Global SFR - HI mass relation for two GASP jellyfish galaxies (JO201 and JO206), compared to a sample of normal spirals of similar masses (grey distribution). In jellyfish galaxies there is a clear excess of SF for their HI mass. From Ramatsoku et al. (2020), see also Deb et al. (2022).}
   \label{fig1}
\end{center}
\end{figure}

\begin{figure}[b]
% \vspace*{-2.0 cm}
\begin{center}
\centerline{\includegraphics[width=2.7in]{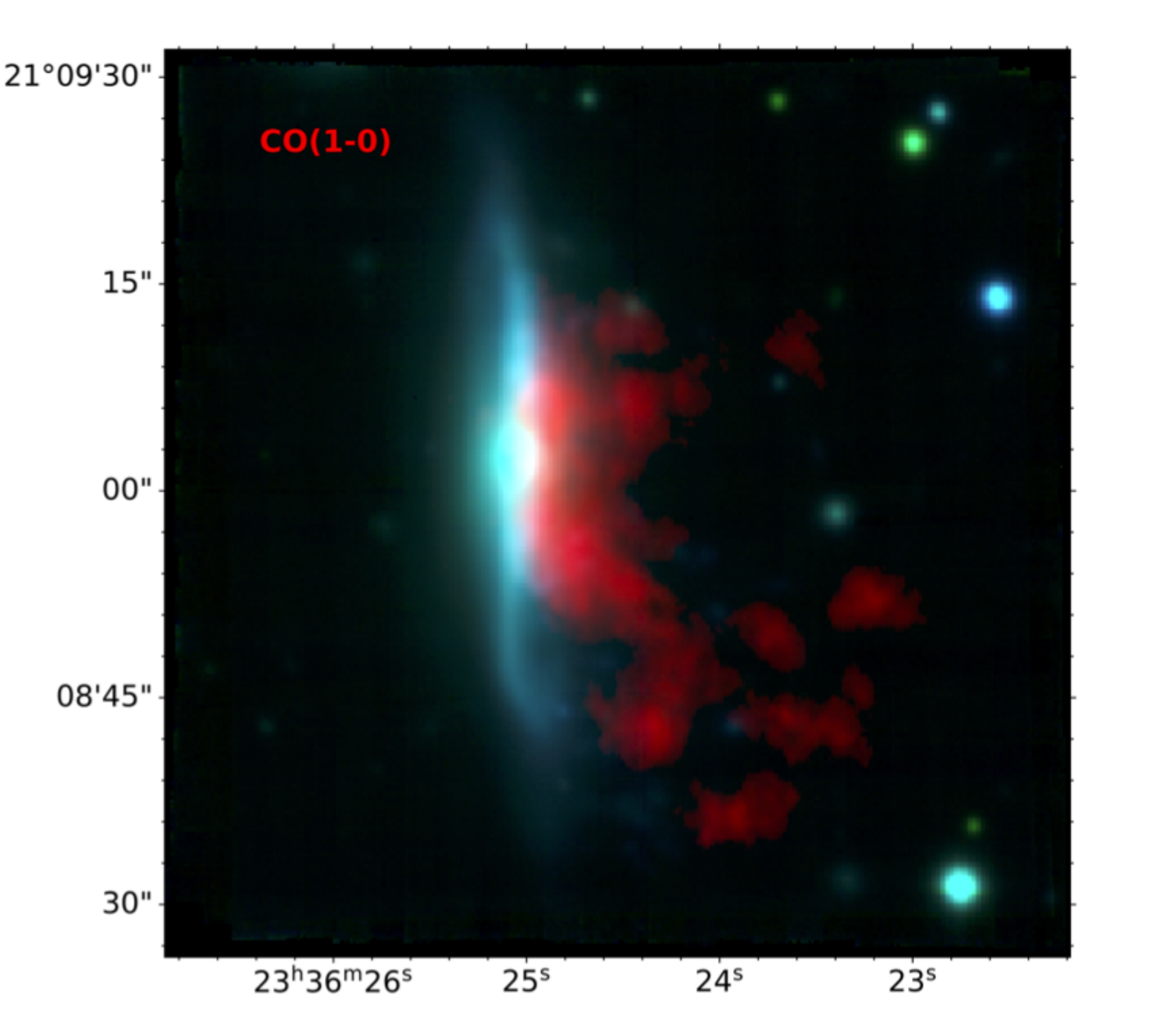}\includegraphics[width=2.7in]{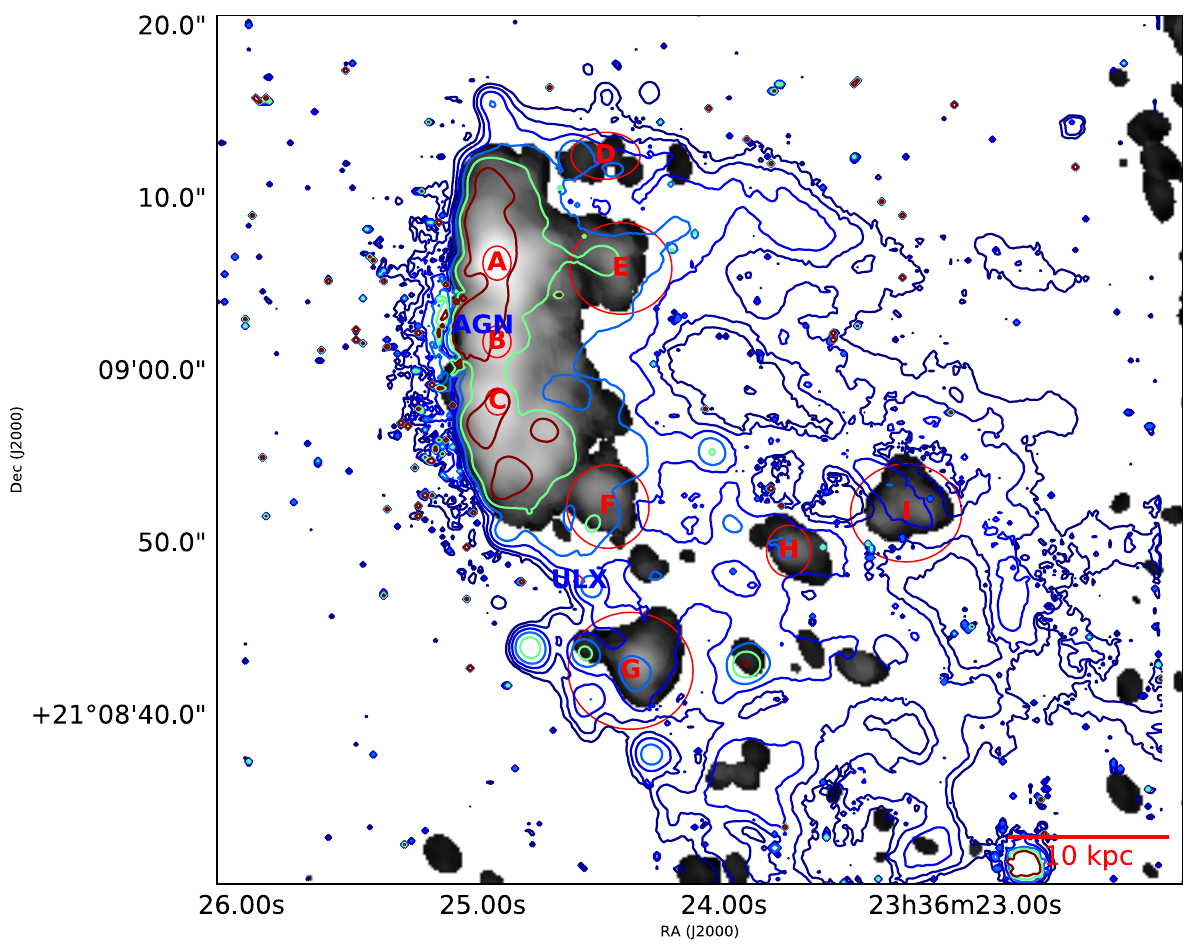}} 
% \vspace*{-1.0 cm}
 \caption{ALMA molecular gas (CO(1-0) emission in the jellyfish galaxy JW100 (red in the left panel, grey in the right panel). Left. There is a large amount of extraplanar CO, in clouds with molecular gas masses ranging from $10^6$ to $10^9 M_{\odot}$. Right. Blue contours are the $\rm H\alpha$ emission. While molecular gas close to the disk may be partly stripped, the clouds far out in the tail must have formed in-situ. From Moretti et al. (2020b).}
   \label{fig1}
\end{center}
\end{figure}

\begin{figure}[b]
\vspace*{-0.5 cm}
\begin{center}
  \includegraphics[width=3.8in]{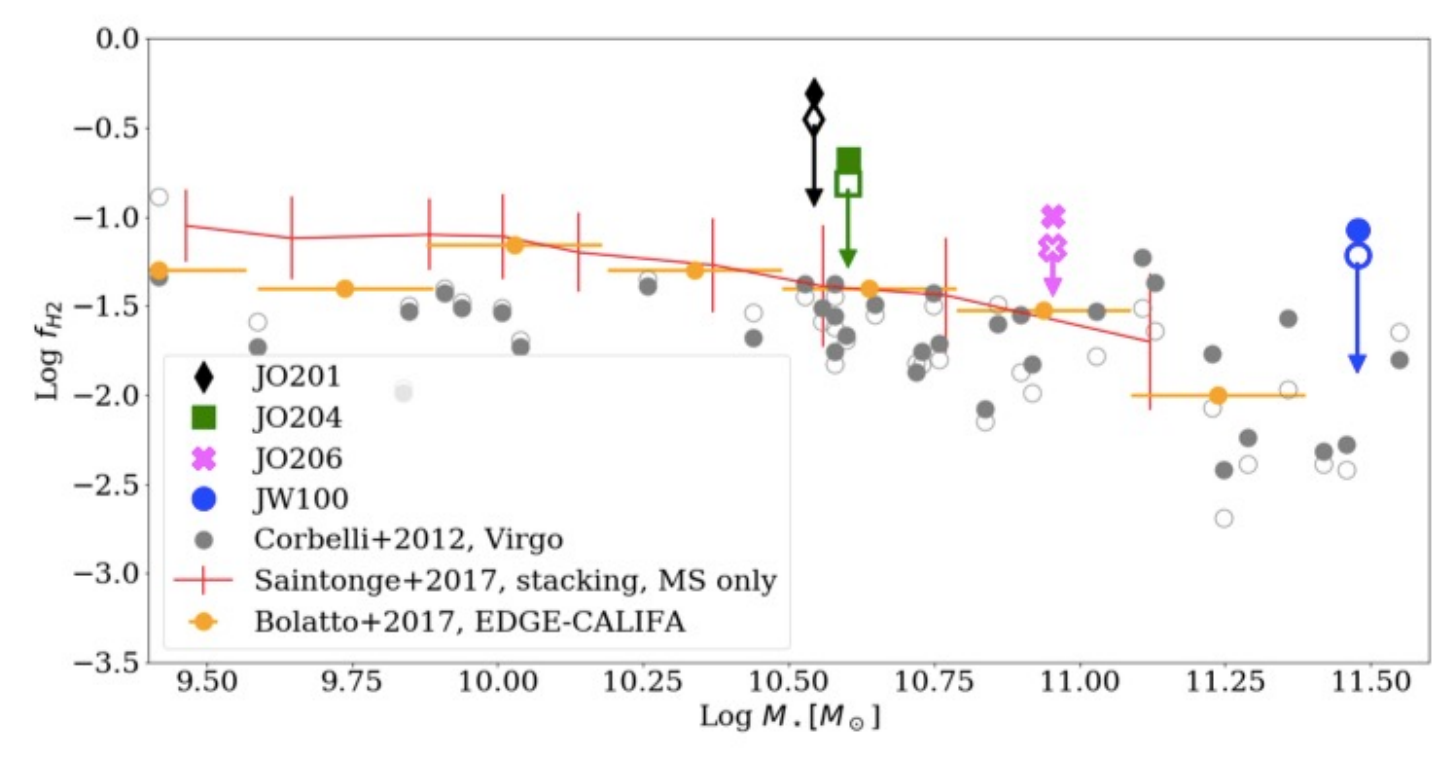}
  \includegraphics[width=4.3in]{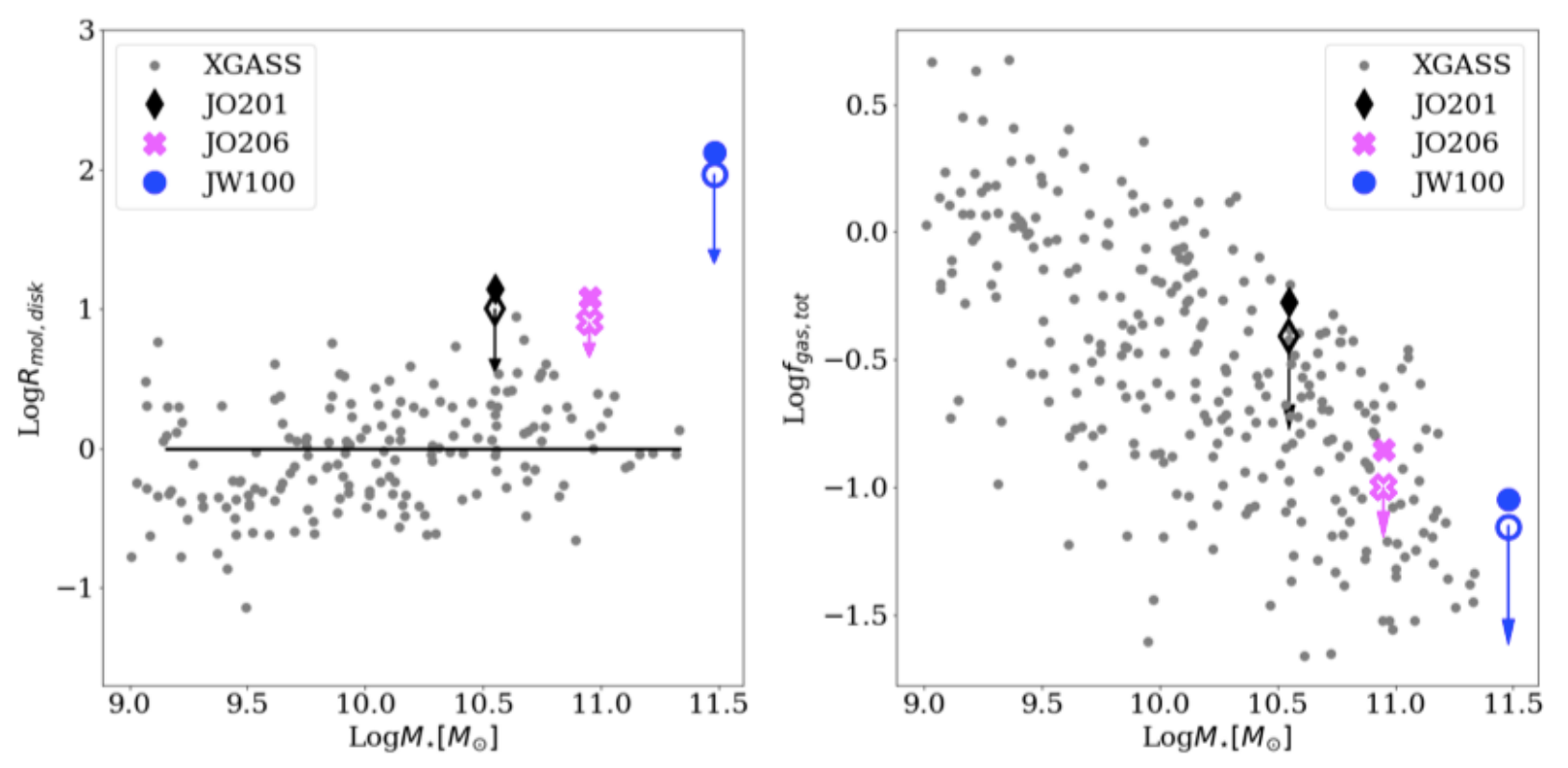}
% \vspace*{-1.0 cm}
 \caption{The molecular gas content of 4 GASP jellyfish galaxies obtained with ALMA (colored symbols) is compared with samples of normal galaxies (grey points and lines). Top. Ratio of molecular gas mass and galaxy stellar mass as a function of stellar mass. Bottom left. Molecular to neutral gas mass ratio versus stellar mass. Bottom right. Total (molecular+neutral) gas mass over stellar mass as a function of stellar mass. From Moretti et al. (2020b).}
   \label{fig1}
\end{center}
\end{figure}

\begin{figure}[b]
%\vspace*{-2.0 cm}
\begin{center}
 \includegraphics[width=4.0in]{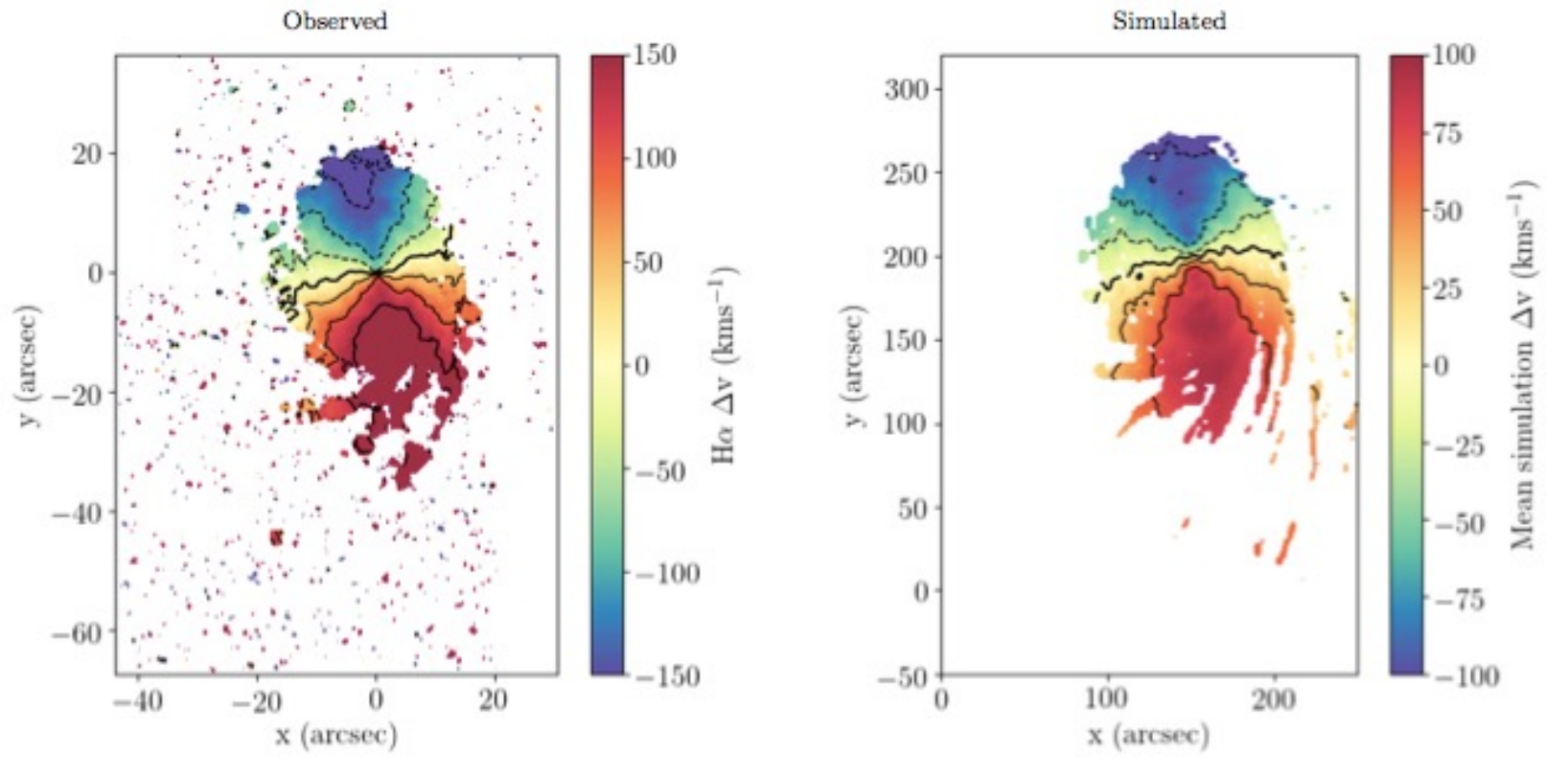}
%\vspace*{-0.4 cm}
 \caption{Example of unwinding arms in a RP stripped galaxy (JO200). The observed $\rm H\alpha$ velocity map (left) is compared with RP simulations (right). From Bellhouse et al. (2021).}
   \label{fig1}
\end{center}
\end{figure}

Neutral gas studies were the first direct evidence for gas stripping
in clusters, revealing HI tails, truncated HI disks, HI disturbed
morphologies and a progressively increasing HI deficiency towards the
central regions of clusters (see Cortese et al. 2021 for a recent
review).  By now, observations of extraplanar tails of stripped material have been obtained
with several different methods (HI line, integral-field spectroscopy,
$\rm H\alpha$ imaging, X-ray emission, radio continuum, UV and optical
imaging, see Boselli et al. 2022), showing tails of gas in different phases and even tails of stars
formed in the stripped gas. It is important to realize that each of
the methods for identifying galaxies
affected by ram pressure provides only a partial view of this phenomenon.
%In this talk I have been asked to focus on integral-field spectroscopic results, and I 

The GASP (GAs Stripping Phenomena in galaxies) project investigates the physical mechanisms that remove gas from galaxies and their effects on the star formation activity and galaxy structure. This programme has been studying galaxies in low redshift clusters, groups, filaments and isolated galaxies, but in these
proceedings I will focus on cluster galaxies only. Readers are refereed to Vulcani et al. (2021) for an overview of non-cluster galaxies.

GASP provides the only large sample of confirmed ram-pressure stripped
galaxies with integral-field (IF) spectroscopy. It consists of 64 ram-pressure
stripped galaxies with a wide range of
stellar masses ($10^9$-$10^{11.5} M_\odot$), that were selected for showing unilateral debris in B-band images. They are hosted in 39
low redshift clusters (z=0.04-0.07) whose velocity dispersions range from
$\sim 500$ to $\sim 1400 \rm km \, s^{-1}$.\footnote{Cases of GASP ram-pressure stripped galaxies in groups
and even in filaments have been presented in Vulcani et al. (2021) and
references therein.}
These galaxies have various stages and degrees of stripping, from weak
or initial signs of stripping, to very strong stripping (``jellyfish
galaxies'', with long gas tails, Fig.~1), to the final stages of
stripping (truncated disks with gas only left in the galaxy
center). Their properties can be contrasted with the GASP control
sample of 30 undisturbed galaxies.  Recently, MUSE data of distant
galaxy clusters have started to allow the first detailed studies of ram-pressure
ionized gas tails also at higher redshifts (2 galaxies at z=0.7, Boselli et al. 2019; 13 galaxies at z=0.3-0.4, Moretti
et al. 2022, Bellhouse et al. 2022).

The GASP survey is based on a MUSE ESO Large Program (spatial resolution $\sim 1 $kpc, galaxy coverage out to 7 effective radii on average), complemented by multi-wavelength follow-up
programs with ALMA, APEX, JVLA, MeerKAT, ATCA, UVIT@ASTROSAT, LOFAR
and Chandra.  Thus, the GASP multi-wavelength coverage allows us to
investigate all the main processes related to star formation, including gas
in different phases, the stellar content and non-thermal processes.

%\begin{figure}[b]
%% \vspace*{-2.0 cm}
%\begin{center}
% \includegraphics[width=3.4in]{Path.eps} 
%% \vspace*{-1.0 cm}
% \caption{Path of pre-solar grains from their stellar sources to the laboratory.}
%   \label{fig1}
%\end{center}
%\end{figure}

\section{Global and spatially resolved star formation rate-mass relation in the disks of galaxies}

The existence of a relation between the integrated star formation rate (SFR) and the total
stellar mass of galaxies (hereafter, ``global'' SFR-M relation) has
been known for many years and has been observed from z=0 to beyond
z=2.  A correlation between the stellar mass surface density and the
SFR surface density has been found in all integral-field spectroscopy
surveys, even at higher redshifts, suggesting that the global relation
originates from a perhaps more fundamental relation on small (1kpc)
scales (Wuyts et al. 2013, Sanchez et al. 2013).
%, Hall et al. 2018,
%Error-Ferrer et al. 2019). 
The scatter in this relation is related to
Hubble type (Gonzalez-Delgado et al. 2016) and variations in star
formation efficienty (Ellison et al. 2020). Reaching larger
galactocentric radii than any other large IF survey, GASP has shown
that the resolved SFR-M correlation of undisturbed galaxies
is broad, and that the scatter
mainly arises from bright off-center star-forming knots. Moreover,
each galaxy has a distinct resolved relation and the global relation
appears to be driven by the existence of the size-mass relation
(Vulcani et al. 2019).

Ram-pressure stripped galaxies show a moderate enhancement of SFR in the disk
for their stellar mass, lying on average above the global SFR-M relation
of undisturbed galaxies by 0.2 dex (Fig.~2, Vulcani et al. 2018,
Roberts \& Parker 2020). Thus, during the initial and strongest stages of
stripping, ram pressure enhances SF before leading to quenching. On
spatially resolved scales, we find again an enhancement of
SFR density at a given mass density at all galactocentric distances, 
consistent with being induced by RP compression waves (Vulcani et al. 2020b).

\section{From stripping to quenching: post-starburst galaxies and AGN}

RP stripping of the disk, and consequent quenching, proceeds mostly outside-in, or
side-to-side if the galaxy is plunging edge-on into the ICM. It is
therefore common to observe RP stripped galaxies still with ionized gas and intense star formation in their central regions as well as long tails of
ionized gas, but having their outer disk regions already devoid of
ionized gas and recently quenched (Gullieuszik et al. 2017, Poggianti
et al. 2017, 2019b, Bellhouse et al. 2017, 2019, Werle et al. 2022). These
outer regions present typical post-starburst (PSB) spectra (Dressler \& Gunn 1983, Poggianti et al. 1999), with no
emission lines and strong Balmer lines in absorption, indicative of a
local sudden truncation of the star formation activity at some point
during the past $\sim 0.5$Gyr. It has been suggested that the strong decline of star formation is responsible for the global radio continuum excess of RP stripped galaxies compared to their ongoing SFR seen by LOFAR
(Ignesti et al. 2022a,b), and that the high radio-to-$\rm H\alpha$ ratio in tails compared to disks is due to relativistic electrons advected from the disks.

If gas is totally stripped, the end product of this evolution is a PSB
galaxy, with no ionized gas and no SF left anywhere in the disk. The
GASP samples, both at low-z and in distant clusters, include by construction several
PSB galaxies whose quenching history can be studied in details applying
spectrophotometric codes (Fritz et al. 2017) to the MUSE spectroscopic
datacube (Vulcani et al. 2020, Werle et al. 2022). The stellar history
maps confirm indeed mostly outside-in and side-to-side quenching, with
characteristic signatures indicating that most cluster PSBs originate
from RP stripping (Vulcani et al. 2020, Werle et al. 2022). The MUSE
spectra also reveal that generally, a strong local starburst takes
place before the SF truncation, accounting for significant fractions
of the mass formed (Fig.~3). 
Moreover, the quenching
timescales and the total time it takes to quench all spaxels can be derived (Werle et al. 2022).

Interestingly, a small fraction of the PSB galaxies in distant
clusters present an inside-out quenching, associated with AGN episodes
which are still detectable from the MUSE emission lines in the centers (Werle et al. 2022). The quenching
history in these cases appears to be dominated by AGN feedback, though this
may be indirectly induced by RP. In fact, an usually high incidence of AGN has been found among strongly RP stripped galaxies (Poggianti et al. 2017), though not all studies find an AGN excess (Roman-Oliveira et al. 2019). In some cases these AGNs show outflows of ionized gas (Radovich et al. 2019) and strong AGN feedback sweeping a large region around the galaxy center (George et al. 2019).

Recently, based on a combination of GASP galaxies and all available literature data for RP stripped galaxies, contrasted with a MANGA mass-matched undisturbed sample, Peluso et al. (2022) has confirmed that the incidence of AGN is higher in RP stripped galaxies. This phenomenon seems to occur preferably during the strongest phase of stripping, when the gas tails are longest. Hydrodynamical simulations reach similar conclusions (Ricarte et al. 2020, Farber et al. 2022). Simulations have previously shown that loss of angular momentum due to the interaction of the rotating ISM with the non-rotating ICM can potentially draw gas on lower orbits, and recent work has clarified the respective roles of mixing and torques from pressure gradients (Akerman et al. 2022 submitted).

\section{Star formation in the tails of stripped gas and gas mixing with the intracluster medium}

The fact that new stars can form in-situ in the stripped gas, outside
of the galaxy disk and even far out in the stripped tails, has been
shown by many studies using several different SF indicators (e.g. Yagi et
al. 2007, Smith et al. 2010, Sun et al. 2010, Merluzzi et al. 2013, Fumagalli et al. 2014,
Fossati et al. 2016, Consolandi et al. 2017, Boselli et al. 2018,
Abramson et al. 2011, Kenney et al. 2014, George et al. 2018, Cramer et al. 2019).

In the extraplanar $\rm H\alpha$ emitting tails of GASP stripped
galaxies the dominant ionization mechanism is indeed photoionization
by young massive stars, as obtained by MUSE BPT diagrams (Poggianti et
al. 2019a). This SF takes place in $\rm H\alpha$-bright, dynamically
cold star-forming clumps formed in-situ in the tails with luminosities
similar to giant and super-giant HII regions (Fig.~1, Poggianti et al. 2019a).

The star formation occurring in the tail can produce from a negligible
fraction to up to $\sim 20\%$ of the total SFR of the system (disk+tail)
(Poggianti et al. 2019a), and the fraction of star formation in the
tails roughly follows the fraction of gas that is stripped according
to traditional analytical formulation (Gullieuszik et al. 2020): the
SFR in the tail can thus be roughly predicted, in a statistical sense,
knowing four main observable quantities: galaxy mass, cluster mass,
galaxy line-of-sight velocity within the cluster and projected
clustercentric distance.

With HST, using broad band filters spanning from UV to I-band and a
narrow-band filter covering $\rm H\alpha$, we can now study at higher
spatial resolution (70pc) the star-forming clumps, in the disks, in
the vicinity of the disk but ``extraplanar'', and in the
tails. For a large sample of both $\rm H\alpha$-selected (over 2400
clumps) and UV-selected (over 3700) clumps in GASP jellyfish galaxies we have been able to study
sizes, luminosities, hierarchical structure as well as stellar masses,
star formation histories and stellar ages (Giunchi et al. submitted,
Werle et al. in prep.). $\rm H\alpha$ and UV luminosity and size
distribution functions, together with luminosities-size relations, can
be found in Giunchi et al., (2022, submitted).

The emerging picture
is that basic star-forming clump properties such as
luminosity and size do not depend strongly on whether the clumps are
sitting in a galaxy disk or far out in the tail without an underlying
disk. These characteristics of GASP clumps resemble
more closely those in highly turbulent, $\rm H\alpha$-bright galaxies
(Fisher et al. 2017) than those in normal low-z spiral galaxies
(see Giunchi et al. 2022 for details).

Interestingly, $\rm H\alpha$ and UV-selected clumps can be embedded in
larger star-forming complexes emitting at optical wavelengths
($V$-band), whose total masses range typically from $10^4$ to $10^7
M_{\odot}$ (Werle et al. in prep.). Thanks to these data, it is now
possible to investigate the fate of these clumps, whether they are
likely to remain intracluster isolated entities (Globular clusters?
Ultracompact dwarf galaxies? Ultradiffuse galaxies?), or get dispersed
and go to contribute to the general intracluster light.

Stripped material and intracluster material can exchange baryons in both directions.
Evidence for mixing between the stripped interstellar medium (ISM) and the
intracluster medium (ICM) has accumulated over the past few years, based
e.g. on X-ray (Poggianti et al. 2019b, Campitiello et al. 2020, Sun et
al. 2022, Bartolini et al. 2022) and on the analysis of the diffuse ionized gas (Tomicic et
al. 2021a,b).  The most direct evidence for ICM-ISM mixing has come
from the gas metallicity in the stripped gas, which decreases along
the tails, going away from the disk (Fig.~4, Franchetto et al. 2021), as
predicted also in hydrodynamical simulations (Tonnesen \& Bryan 2021).

Overall, the data supports a scenario in which the stripped gas gets
mixed with the ICM, accreting a significant fraction of mass, and still manages to cool and collapse to form new
stars. Observations of the magnetic field aligned with the tail in the
stripping direction in the JO206 GASP jellyfish galaxy favor a
``magnetic draping'' scenario (the galaxy moving within the
magnetized ICM sweeping up the surrounding magnetic field, Dursi \& Pfrommer 2008) and point
out that the magnetic field may play a role favoring the SF in the tails by reducing the thermal conduction between ICM and ISM, hence the ISM evaporation
(Mueller et al. 2020).

In order to complete the picture, there is another crucial stage
in the star formation process: the molecular cloud formation from the
neutral gas, as discussed below.

\section{Neutral gas, molecular gas and star formation}

As mentioned above, no selection method can provide the whole census
of ram pressure stripped galaxies. It is thus of fundamental
importance to obtain multi-wavelenght information for samples selected
in different ways. Only then we will be able to understand under what
conditions multiphase tails are formed, and what is the evolutionary
sequence among the observations of tails seen at different
wavelengths. There are still relatively few RPS galaxies for which a
detailed multi-lambda coverage is available (e.g. ESO137-001 and JW100).
%Poggianti et al. 2019 and Deb et al. 2022). 

For a subset of the GASP jellyfish galaxies, both HI and CO
observations have been collected, in addition to MUSE.  These GASP
jellyfish galaxies are slightly HI deficient compared to similar
non-stripped galaxies but have a significant excess of star
formation for their HI content (Fig.~5, Ramatsoku et al. 2019,
2020, Deb et al. 2020, 2022, see also Luber et al. 2022 for an HI comparison of jellyfish and other cluster galaxies).

These same galaxies have very large amounts of molecular gas, as
estimated from their CO(2-1) and CO(1-0) emission (Fig.~6, Jachym et al. 2017, 2019, Moretti et
al. 2018b, 2020a,b).  Their molecular gas mass at a given stellar mass is
4-5 times higher than in undisturbed galaxies (Fig.~7). The molecular to
neutral gas ratio in their disks is between 4 and 100 times higher
than normal. Surprisingly, overall, the total (molecular plus neutral)
gas mass is similar to that in normal galaxies. These results, shown
in Moretti et al. 2020b, are solid irrespective of the conversion
factors used for CO to $H_2$ and they suggest a very efficient conversion
of neutral gas into molecular gas in jellyfish galaxies.

As fas as the star formation efficiency is concerned (ratio between
star formation rate and molecular gas mass) this appears to be significantly
lower in the tails 
%(and sometimes also in the disk) 
than in normal spirals, with long depletion times up to $10^{10}$ yr
(Moretti et al. 2020b, and in prep.).

MeerKAT is soon to provide HI-selected samples for which additional
multi-wavelength informations, including molecular gas estimates, have been secured.
%(Serra et al. in prep., Moretti et al. in prep.).

\vspace{-0.7cm}
\section{Conclusions}
Ram-pressure stripped galaxies are unique laboratories to study the star
formation process and the baryonic cycle under unusual conditions. Gas clouds in stripped tails are embedded within the hot ICM and do not experience the influence of an underlying stellar disk, yet stars are commonly formed in stripped tails of multi-phase gas. Also in the galaxy disks, SF appears to be slightly enhanced globally, and strongly enhanced locally at any given time, where gas is still left. We witness an exceptionally efficient conversion of neutral to molecular gas in these galaxies, while the star formation efficiency ranges from usually normal in disks to much lower than normal in tails. PSB galaxies are the natural end-product of the stripping and subsequent quenching, and AGN activity possibly triggered by RP itself sometimes contributes in an inside-out manner to the quenching.
I have presented only some of the highlights of the GASP project concerning the star formation activity and the baryonic cycle in cluster galaxies. I have not presented the consequences of the RP-induced SF on the galaxy structure: as an example, RP stripping can cause the unwinding of spiral arms without the contribution of tidal interactions 
(Fig.~8, Bellhouse et al. 2021).
Interested readers can find the full list of GASP publications at
https://web.oapd.inaf.it/gasp/.

\acknowledgements
Based on observations collected at the European Organization for
Astronomical Research in the Southern Hemisphere under ESO programme
196.B-0578. This project has received funding from the European
Research Council (ERC) under the European Union's Horizon 2020
research and innovation programme (grant agreement No. 833824).

%\begin{discussion}
%\end{discussion}

\end{document}